\documentclass[12pt,a4paper]{article}
\usepackage{amsmath, amssymb, fullpage, graphics}
\usepackage{caption, graphicx, fancyhdr, fancybox}
\usepackage[cp866]{inputenc}
\usepackage{color}

\begin{document}

\begin{center}
{\sf STABILITY OF SHEAR SHALLOW WATER FLOWS WITH FREE SURFACE} \\
\vspace{2mm}

A.~A.~Chesnokov$^{1}$, G.~A.~El$^{2}$, S.~L.~Gavrilyuk$^{3}$, and M.~V.~Pavlov${^1}$ \\[2mm]

${^1}$Novosibirsk State University, \\
1 Pirogov Str. Novosibirsk, 630090, Russia \\[2mm]
${^2}$Department of Mathematical Sciences, Loughborough University, \\ 
Loughborough LE11 3TU, United Kingdom \\[2mm]
${^3}$Aix-Marseille Universit\'{e}, UMR CNRS 7343, IUSTI, \\
5 rue E. Fermi, 13453 Marseille CEDEX 13, France \\[2mm]

{\sf e-mails:} chesnokov@hydro.nsc.ru, g.el@lboro.ac.uk, \\ sergey.gavrilyuk@univ-amu.fr, mpavlov@itp.ac.ru
\end{center}

\begin{abstract}
Stability of inviscid shear shallow water flows with free surface is studied in the framework of the Benney equations. This is done by investigating the generalized hyperbolicity of the integrodifferential Benney system of equations. It is shown that all shear flows having monotonic convex velocity profiles are stable. The hydrodynamic approximations of the model corresponding to the classes of flows with piecewise linear continuous and discontinuous velocity profiles are derived and studied. It is shown that these approximations possess Hamiltonian structure and a complete system of Riemann invariants, which are found in an explicit form. Sufficient conditions for hyperbolicity of the governing equations for such multilayer flows are formulated. The generalization of the above results to the case of stratified fluid is less obvious, however, it is established that vorticity has a stabilizing effect. 
\end{abstract}


\section{Introduction}
The classical shallow water equations~\cite{Stoker} describe the propagation of long waves on a free surface under the assumption that the flow under consideration is potential. In this model, only the averaged over depth velocities are used in the formulation of the governing equations. However, in practice, fluid flows are sheared, which is mainly due to viscosity effects near boundaries. Obviously, for a more accurate modelling of the wave propagation it is necessary to also take into account non-uniformity of the flow.

An extension of the classical shallow water theory to a  vortical fluid flow was proposed by Burns~\cite{Burns}, who was the first to study a plane parallel shear flow of inviscid fluid in linear approximation and derive the dispersion relation for normal modes. In this case, the speed of perturbation propagation is determined by an integral relation depending on the horizontal velocity profile over depth. A nonlinear model of long surface waves in shear flow was derived by Benney~\cite{Benney} and represents an integrodifferential system of equations in sharp contrast with the quasilinear system for potential flow. Nevertheless, it was shown in \cite{Benney} that the Benney system can be written in the form of the so called infinite-component moment chain  and, similar to the classical shallow water equations, possesses infinitely many conservation laws. Zakharov~\cite{Zakharov} established equivalence between the Benney system and the Vlasov kinetic equation, and Teshukov et al.~\cite{TRC} found an explicit transformation between these two models. Kupershmidt and Manin~\cite{KupMan}, Lebedev and Manin \cite{LebMan} found local Hamiltonian structure and the Lax pair respectively for the Benney system. Families of exact solutions, having the structure of travelling and simple waves, were constructed and interpreted by Freeman~\cite{Freeman}, Sachdev~\cite{Sachdev}, Varley and Blythe~\cite{Varley} and Teshukov et al.~\cite{TRC}. 

Zakharov~\cite{Zakharov} considered the first non-trivial multi-component reduction of the Benney system for multi-layered fluid, proved its integrability  and constructed a complete infinite set of conservation laws for this reduction. Zakharov's multilayer reduction has the following important property: at each point $x$ the horizontal  components of the velocity within each layer are constant and distinct. This property prevents the applicability of Zakharov's reduction to the description of actual multilayer shallow water flows since sliding between layers is unusual. 

Stability of shear flows for the full Euler equations is a fundamental problem of Fluid Mechanics (see e.g.~\cite{Drazin}). The classical stability and instability criteria  formulated in terms of growth of linear perturbations (Rayleigh, Fjortoft) are usually obtained for flows between rigid walls. Some recent works use the generalized notion of stability as the well-posedness of time evolution, i.e. hyperbolicity  (see~\cite{Chumakova_etal1, Chumakova_etal2}), but they also consider either flows under closed lid or use periodic boundary conditions in the vertical direction  which greatly simplifies the analysis. However, the presence of  free surface can obviously change the flow stability criteria and, to our knowledge, stability of shallow water shear flows with a free surface has not been studied before.  We note that, being an integrodifferential system, the Benney equations cannot be directly classified in terms of hyperbolicity. A generalized theory of characteristics and the notion of hyperbolicity for integrodifferential equations of the long wave theory was introduced by Teshukov~\cite{Tesh85, Tesh94, LT00}. Recently Chesnokov and Khe~\cite{Chesn13} revealed an analogue of Landau damping for the Benney equations.

The structure of the paper is as follows. In Section 2, we present the necessary preliminaries on the hyperbolicity of the Benney equations in sense of \cite{Tesh85,  Tesh94}. We then show that, if the velocity profile over vertical coordinate is smooth, then monotonicity and convexity of the velocity profile is sufficient for the stability of flows with a free surface. We also extend this result for the Fjortoft-like velocity profiles.  Using the Vlasov-like formulation of the governing equations we reveal analogy between the criteria of the stability of the plasma waves and shear flows. In Section 3 we derive the models corresponding to the class of flows with a piecewise linear (continuous or discontinuous) velocity profile. We then formulate sufficient conditions for the stability of such multilayer flows. The study of stability is based on the verification of the hyperbolicity condition for the governing equations. We also reveal some important mathematical properties of the equations of multilayer flows (the existence of Riemann invariants and the Hamiltonian structure, which implies integrability). It should be stressed that our approximation of the Benney equations for shear flows with piecewise constant vorticity is an important ``upgrade'' of the classical Zakharov reduction~\cite{Zakharov} as it admits a class of physically natural continuous velocity profiles. In Section 4, we consider two-layer stratified flows with a piecewise linear velocity profile. We show that the generalization of previous results to the case of stratified flows is hardly possible. Nevertheless, we can state that the presence of vorticity has a stabilizing effect on the flow of stratified fluid. Finally, we draw conclusions from our study. 

\section{Benney equations and the hyperbolicity condition} 
The system of Benney equations~\cite{Benney}
\begin{equation} \label{eq:VSW}
  \begin{array}{l}\displaystyle
    u_t+uu_x+wu_z+gh_x=0, \\[3mm]\displaystyle
    h_t+\bigg(\int_0^h u\,dz\bigg)_x=0, \quad w=-\int_0^z u_x(t,x,z')\,dz'
  \end{array}
\end{equation}
describes the propagation of nonlinear long waves in a shear flow of an ideal homogeneous fluid layer with a free boundary $z = h(t,x)$ over a flat bottom $z = 0$ under gravity field. Here $t$ is the time, $x$ and $z$ are the Cartesian coordinates, $u(t,x,z)$ and $w(t,x,z)$ are the horizontal and vertical components of the velocity vector respectively, $g$ is the acceleration due to gravity. From Eqs.~\eqref{eq:VSW} one can deduce that the long-wave vorticity $\omega=u_z$ is conserved along the trajectories: 
\begin{equation}\label{eq:vort_eq}
  \omega_t+u\omega_x+w\omega_z=0. 
\end{equation}

Generalized hyperbolicity conditions for the integrodifferential equations~\eqref{eq:VSW} on a solution $u(t,x,z)$, $h(t,x)$ with a monotonic velocity profile (e.g. $u_z>0$) are formulated in \cite{Tesh85, LT00, TRC} in terms of the characteristic function
\begin{equation} \label{eq:chi}
	\chi(k)= 1 - g\int_0^h \frac{dz}{(u-k)^2}= 1 + g\int_0^h \frac{1}{u_z} 
	\frac{\partial}{\partial z}\bigg(\frac{1}{u-k}\bigg)\,dz
\end{equation}
or, more precisely, in terms of its limit values on the interval $[u_b, u_s]$ (where the subscripts $b$ and $s$ correspond to values of the functions at $z=0$ and $z=h$ respectively) from the upper and the lower complex half-planes
\begin{equation} \label{eq:chi-pm}	
  \chi^\pm(u)=1 + g\bigg( \frac{W_s}{u_s-u} - \frac{W_b}{u_b-u} -
  \int_0^h \frac{W'_z\,dz'}{u'-u} \mp \pi i \frac{W_z}{u_z} \bigg)\,, 
\end{equation}
which are obtained from~\eqref{eq:chi} by integration by parts and the application of the Sokhotski--Plemelj formulae. Here $W=1/u_z$,  $u'=u(t,x,z')$, $W'=W(t,x,z')$.

\begin{figure}[tbp]
\begin{center}
\resizebox{1\textwidth}{!}{\includegraphics{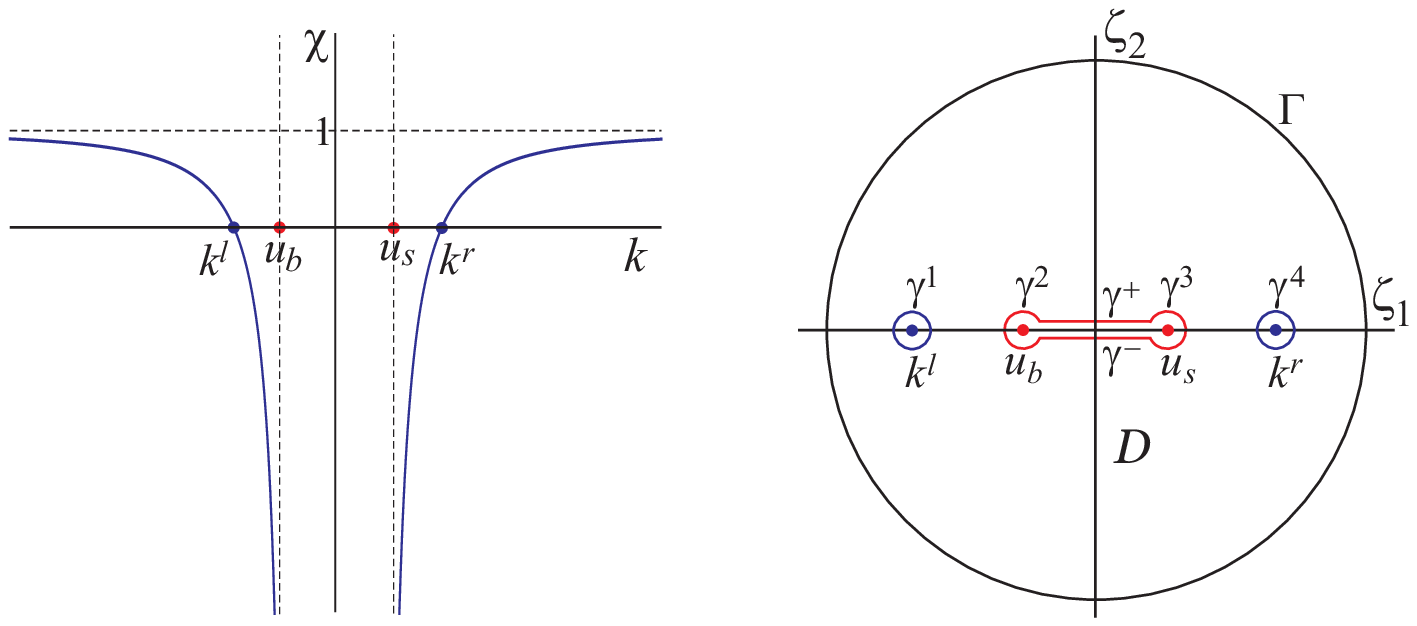}} \\[0pt]
\end{center}
\parbox{0.48\textwidth}{\caption{A typical graph of the function $\chi(k)$, $k\in (-\infty, u_b) \, \cup \, (\infty,u_s)$ for monotonic ($u_z>0$) velocity profile.} \label{fig:chi}} \hfill
\parbox{0.48\textwidth}{\caption{Contour in the complex plane $\zeta$ used for the formulation of the hyperbolicity conditions.} \label{fig:contour}}
\end{figure}

Let the bounded function $W>0$ be differentiable with respect to $z$, and $W_z$ is H\"older continuous on the interval $z \in [0,h]$. Then the characteristic equation $\chi(k)=0$ has exactly two real roots $k=k^l<u_b$ and $k=k^r>u_s$ (see Figure~\ref{fig:chi}). Indeed, $\chi(k)\to 1$ if $k\to\pm\infty$; $\chi(k)\to -\infty$ if $k\to u_b-0$ or $k\to u_s+0$; $\chi'(k)<0$ for $k<u_b$ and $\chi'(k)<0$ for $k>u_s$. Eqs.~\eqref{eq:VSW} are hyperbolic (in the sense of \cite{Tesh85, Tesh94}) if the following condition holds
\begin{equation}\label{eq:hyp-cond}
  \Delta{\rm arg}\,\frac{\chi^+(u)}{\chi^-(u)}=0, \quad \chi^\pm(u)\neq 0.
\end{equation}
The argument increment is calculated when $z$ changes from $0$ to $h$ at fixed values of variables $t$ and $x$.

The condition \eqref{eq:hyp-cond} provides, in particular, the absence of complex roots of the characteristic equation $\chi(k)=0$. Consider the domain $D$ (see Figure~\ref{fig:contour}) in the plane of complex variable $\zeta=\zeta_1+i\zeta_2$, bounded by a circle $\Gamma$ of radius $R_\varepsilon$ with centre at the origin of coordinates, circles $\gamma^j$ of radii $r_\varepsilon$ with centres at the points $k^l$, $k^r$, $u_b$, $u_s$ and segments $\gamma^\pm$ of the cut sides $(u_b,u_s)$. One supposes that $r_\varepsilon\to 0$, $R_\varepsilon \to \infty$ as $\varepsilon\to 0$. The increment in the argument of the function $\chi(\zeta)$ along the boundary of the domain $D$ normalized by $2\pi$ is equal to the number of zeros of the function $\chi(\zeta)$ in this domain. Indeed, $\chi(\zeta)$ has no poles in the domain $D$. Moreover, $\chi(\zeta)$ has first-order zeros at the points $\zeta=k^l$, $\zeta=k^r$ and first-order poles at the points $\zeta=u_b$, $\zeta=u_s$. Thus, $\chi(\zeta)$ has no zeros in the domain $D$ if the increment of its argument along segments $\gamma^\pm$ is equal to zero. 

Following \cite{Tesh85, LT00, TRC} we introduce the Riemann invariants
\begin{equation} \label{eq:Riemann-inv} 
   R=u-g\int_0^h \frac{dz'}{u'-u}, \quad
   r^i=k^i-g\int_0^h \frac{dz}{u-k^i} \quad (i=l,r). 
\end{equation}
(Note that equation~\eqref{eq:vort_eq} is already in the Riemann form with $\omega=u_z$ being the Riemann invariant).
Here $k^l$ and $k^r$ are the roots of the characteristic equation $\chi(k)=0$. Let the functions $u(t,x,z)$ and $h(t,x)$ be a solution of Eqs.~\eqref{eq:VSW}, then the Riemann invariants satisfy the equations 
\begin{equation} \label{eq:Riemann-gen}
  \begin{array}{l}\displaystyle
	R_t+uR_x+wR_z=0, \quad \omega_t+u\omega_x+w\omega_z=0, \\[3mm]\displaystyle
	r^l_t+k^l r^l_x=0, \quad r^r_t+k^r r^r_x=0.
  \end{array}
\end{equation}
According to \cite{Tesh85, LT00} systems~\eqref{eq:VSW} and \eqref{eq:Riemann-gen} are equivalent on smooth solutions if the hyperbolicity condition~\eqref{eq:hyp-cond} holds. 
\vspace{1mm}

{\sf Remark 1.} Characteristic properties of Eqs.~\eqref{eq:VSW} for flows with a non-monotonic velocity profile were studied in \cite{Tesh95}. Let the function $u(t,x,z)$ satisfy the following conditions 
\begin{equation} \label{eq:nonmonoton} 
 \begin{array}{l}\displaystyle
   u_z>0 \quad {\rm for} \quad z\in [0,z_*(t,x)), \quad 
   u_z<0 \quad {\rm for} \quad z\in (z_*(t,x),h(t,x)] \\[2mm]\displaystyle 
   u_{zz}(t,x,z_*)\neq 0,  \quad u_b=u(t,x,0)<u(t,x,h),
 \end{array}
\end{equation}
i.e. $z=z_*$ is the point of maximum for $u$ as a function of $z$. We define the complex function
\[ \chi_1(\zeta)=(\zeta-u_*)\bigg(1-g\int_0^h \frac{dz}{(u-\zeta)^2}\bigg), \]
where $u_*=u(t,x,z_*)$. According to \cite{Tesh95} the generalized hyperbolicity conditions of  Eqs.~\eqref{eq:VSW} for flows of class \eqref{eq:nonmonoton} are formulated as follows
\[ \frac{1}{\pi}\Delta{\rm arg}\,\frac{\chi_1^+(u)}{\chi_1^-(u)}=-3, \quad 
   \chi^\pm(u)\neq 0. \]
Here $\chi_1^\pm(u)$ are the limiting values of $\chi_1(\zeta)$ from the upper and lower complex half-planes on the segment $[u_b,u_*]$. The argument increment is calculated when $u$ changes from $u_b$ to $u_*$.  

Unfortunately, singularity of $1/u_z$ at the point $z=z_*$ does not allow us to represent the function $\chi_1(\zeta)$ in the form of the Cauchy type integral (as it was done in \eqref{eq:chi} by integration by parts) and define the functions $\chi_1^\pm(u)$. For this reason we restrict here our consideration to flows with a monotonic velocity profile. 

\subsection{Stability analysis}
Let us study stability of shallow shear flows with a free surface in terms of hyperbolicity of the governing equations~\eqref{eq:VSW}. In particular, we show that for smooth flows with a monotonic and convex velocity profile the hyperbolicity condition~\eqref{eq:hyp-cond} is always fulfilled. 

Let $u=U(z)$, $U'(z)>0$ (the variables $t$ and $x$ are fixed). In the verification of the hyperbolicity condition~\eqref{eq:hyp-cond}, it is convenient to use the functions 
\[ \Psi^\pm(U)=m(U)\chi^\pm(U), \quad m(U)=(U_1-U)(U-U_0) \geq 0 \]
which have no poles at the boundary points $U_0=U(0)$ and $U_1=U(h)$. Here the complex functions $\chi^\pm$ are defined by \eqref{eq:chi-pm}. In the plane $(Z_1,Z_2)$ we construct a closed contour $C$ consisting of the contours $C^-$ and $C^+$. The contour $C^-$ is given parametrically by the equations
\[ Z_1={\rm Re}\{\Psi^-(U)\}, \quad Z_2={\rm Im}\{\Psi^-(U)\}. \]
A contour $C^+$, which is symmetric about the $Z_1$ axis to the contour $C^-$, is given by the same equations with the function $\Psi^+(U)$. If the point of $Z_1=0$, $Z_2=0$ lies in the domain bounded by the contour $C$, then the characteristic equation $\chi(k)=0$ has complex roots (the function $\chi(k)=0$ is given by \eqref{eq:chi}). Otherwise, the governing equations~\eqref{eq:VSW} for the corresponding solution are hyperbolic. 

Taking into account the identity  $W_z=(1/U')'=-U''/(U')^2$ one obtains 
\begin{multline*}
     \Psi^\pm(U)= m(U)\bigg(1+g\int_0^h 
     \frac{U''(\xi)\,d\xi}{(U'(\xi))^2(U(\xi)-U(z))}\bigg) + \\[3mm]\displaystyle
     \quad\quad\quad\quad\quad\quad\quad\quad  
     + g\bigg(\frac{U-U_0}{U_1'}+\frac{U_1-U}{U_0'}\bigg) 
     \pm g \pi i\, \frac{m(U)U''}{(U')^3}\,.
\end{multline*}
We will use these functions to prove the following main statements. 
\vspace{1mm}

{\sf Lemma 1.} {\it Let $U''(z)\neq 0$, then the flow is stable (Rayleigh-like criterion).}
\vspace{1mm}

{\sf Lemma 2.}
{\it Let $U''<0$ for $z \in [0,z_c)$, $U''(z_c)=0$, and $U''>0$ for $z \in (z_c,h]$. Then the flow is stable.}
\vspace{1mm}

{\sf Lemma 3.}
{\it Let $U''>0$ for $z \in [0,z_c)$, $U''(z_c)=0$, and $U''<0$ for $z \in (z_c,h]$. Then the flow may be unstable.}
\vspace{1mm}

We present the proof of these statements.
\vspace{1mm}

{\sf Proof of Lemma 1.} If $U''(z)\neq 0$, then the functions ${\rm Im}\, \Psi^\pm(U)$ have a constant sign in the interval $(U_0,U_1)$ because $U'>0$, $m(U)>0$, and $U''\neq 0$. For $U=U_0$ and $U=U_1$ the imaginary part of $\Psi^\pm(U)$ vanishes  and the functions take the following values at these points: 
\begin{equation} \label{eq:boundary-points}  
  \Psi^\pm(U_0)=g\frac{U_1-U_0}{U_0'}>0, \quad \Psi^\pm(U_1)=g\frac{U_1-U_0}{U_1'}>0.
\end{equation}

\begin{figure}[tbp]
 \begin{center}
  \resizebox{.9\textwidth}{!}{\includegraphics{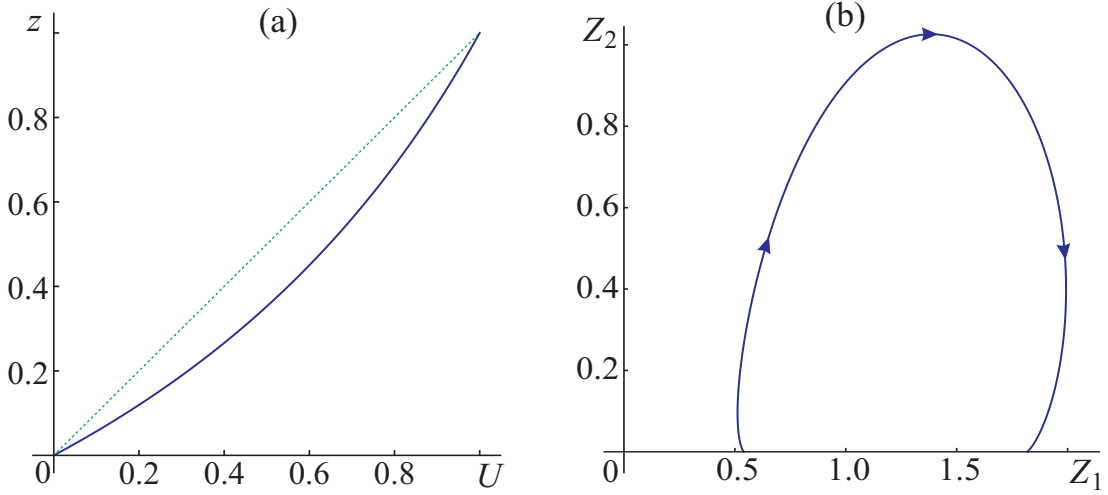}} \\[0pt]
 \end{center}
{\caption{An example of a monotonic convex velocity profile $u=U(z)$ {\rm (a)}  and the corresponding contour $C^-$ in the complex plane $(Z_1,Z_2)$ {\rm (b)} (the arrows indicate the direction of the path tracing).}\label{fig:Rayleigh}} 
\end{figure}

A typical velocity profile $u=U(z)$ and corresponding contour $C^-$ are shown in Figure~\ref{fig:Rayleigh} which is obtained for the function 
\[ U(z)=\frac{(z+1)^{-3/4}-1}{2^{-3/4}-1}, \quad z\in [0,1] \]
with $g=1$ (in this case $U''<0$). The point $Z_1=0$, $Z_2=0$ is not in the domain bounded by the contour $C$. This means that the arguments of the complex functions $\Psi^\pm(U)$ do not increase as $U$ is changed from $U_0$ to $U_1$ and, consequently, the hyperbolicity condition~\eqref{eq:hyp-cond} is satisfied. 
\vspace{1mm}

{\sf Proof of Lemma 2.}
This statement is a {\it Fjortoft-like criterion} which can also be written in the following form. Let 
 \begin{equation} \label{eq:Fjortoft} 
 (U(z)-U_c)U''(z) \geq 0, \quad z\in [0,h]
\end{equation}
then the flow is stable. Here $z=z_c$ is an inflection point at which $U''(z_c)=0$ and $U_c=U(z_c)$. By the definition of the functions $\Psi^\pm$, the sign of ${\rm Im}\Psi^\pm$ coincides with the sign of $U''$. As before, at the boundary points $U_0$ and $U_1$ the functions $\Psi^\pm$ take the positive values. Therefore, the question of the satisfaction of the hyperbolicity condition~\eqref{eq:hyp-cond} reduces to validating the inequality $\Psi^\pm(U_c)>0$. When inequality~\eqref{eq:Fjortoft} is satisfied, we have 
\[ \Psi^\pm(U_c)=g\bigg(\frac{U_c-U_0}{U'_1}+\frac{U_1-U_c}{U'_0}\bigg)+ 
   m(U_c)\bigg(1+ g\int_0^h \frac{(U(z)-U_c)U''(z)\,dz}{(U'(z))^2(U(z)-U_c)^2}\bigg)>0, \]
since all the terms of the expression are positive. 

\begin{figure}[tbp]
 \begin{center}
  \resizebox{.9\textwidth}{!}{\includegraphics{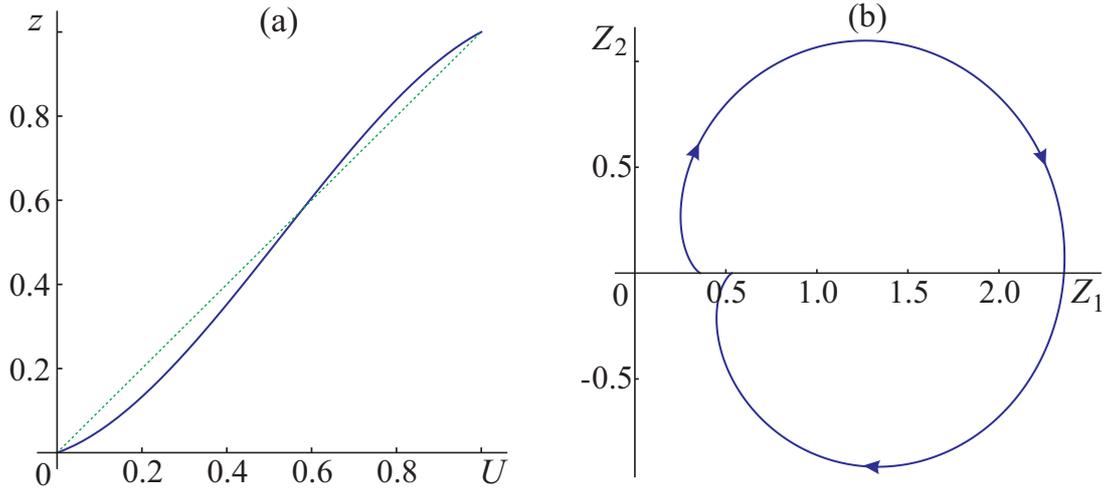}} \\[0pt]
 \end{center}
{\caption{An example of monotonic velocity profile (solid line) which satisfies condition~\eqref{eq:Fjortoft} {\rm (a)} and corresponding contour $C^-$ in the complex plane $(Z_1,Z_2)$ {\rm (b)}.  }\label{fig:Fjortoft}} 
\end{figure}

A typical velocity profile satisfying condition~\eqref{eq:Fjortoft} and corresponding contour $C^-$ are shown in Figure~\ref{fig:Fjortoft}, which is obtained for the function
\[ z(U)=\bigg(\Big(1-\frac{a_1 U}{3}\Big)\frac{a_2 U}{2}+1+\frac{(a_1-3)a_2}{6}\bigg)U, 
   \quad U\in [0,1] \]
(here $U_c=1/a_1$) with $a_1=1.9$, $a_2=3.5$, $g=1$. 
\vspace{1mm}

{\sf Proof of Lemma 3.}
Although at the boundary points inequalities~\eqref{eq:boundary-points} are still satisfied and the imaginary part of the functions $\Psi^\pm$ changes sign once with a change in $U$ from $U_0$ to $U_1$, we can not guarantee that $\Psi^\pm(U_c)>0$. Indeed, let us consider the following class of velocity profiles
\begin{equation} \label{eq:class-st-unst} 
  U(z)=\frac{\tanh((z-z_c)a)+\tanh(a z_c)}{\tanh((1-z_c)a)+\tanh(a z_c)}, 
  \quad z\in [0,1] 
\end{equation}
which corresponds to the considered case. Here $z_c$ is inflection point and $U'''(z_c)<0$. The parameter $a$ affects the rate of change in the function $U(z)$ near the inflection point: the velocity profile tends to a discontinuous piecewise constant function as $a\to \infty$. 

\begin{figure}[tbp]
 \begin{center}
  \resizebox{.9\textwidth}{!}{\includegraphics{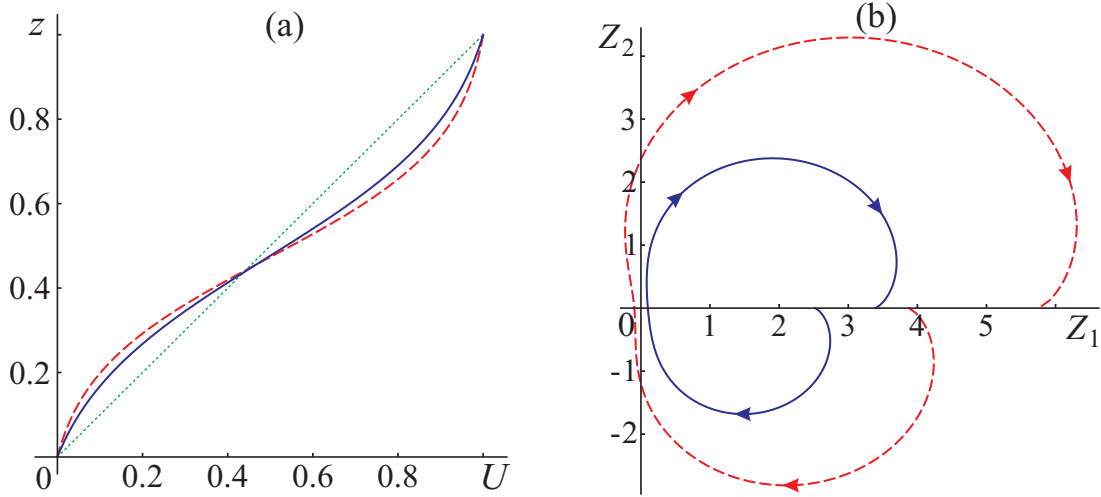}} \\[0pt]
 \end{center}
{\caption{Monotonic velocity profiles from class~\eqref{eq:class-st-unst} {\rm (a)} and corresponding contours $C^-$ {\rm (b)} obtained for $z_c=0.47$, $a=2.8$ (solid lines) and $a=3.5$ (dashed lines). }\label{fig:poss-unst}} 
\end{figure}

Velocity profile of class~\eqref{eq:class-st-unst} and corresponding contour $C^-$ are shown in Figure~\ref{fig:poss-unst} (solids) for $z_c=0.47$, $a=2.8$, and $g=1$. As we can see from Figure~\ref{fig:poss-unst} (b) the point of $Z_1=0$, $Z_2=0$ is not in the domain bounded by the contour $C$. Consequently, the hyperbolicity condition~\eqref{eq:hyp-cond} is fulfilled. Dashed curves in Figure~\ref{fig:poss-unst} are obtained for $a=3.5$ (the others parameters are the same). In this case the point of $Z_1=0$, $Z_2=0$ belongs to the domain bounded by the contour $C$ and the hyperbolicity condition~\eqref{eq:hyp-cond} is violated. 
\vspace{1mm}

We proved that the classical stability criteria for shear flows of ideal fluid~\cite{Drazin} correspond to the hyperbolicity condition~\eqref{eq:hyp-cond} of the governing equations~\eqref{eq:VSW}. Thus the Rayleigh--Fjortoft criteria (for flows with monotonic velocity profile) provide the hyperbolicity of the flow with a free surface, i.e. it is a sufficient condition of the stability for vortex shallow water flows. The same correspondence between the classical stability criteria and the hyperbolicity condition for the integrodifferential equations of the long wave theory for the flow between rigid walls was established in \cite{KnCn12}. 

Let us also remark that the convexity of the velocity profile is not sufficient for the stability, if the dispersive terms are added. For the Serre--Green--Naghdi-type equation the stability criterion was established in \cite{Gavrilyuk_Teshukov} where additional (with respect to the convexity and monotonicity conditions) inequalities were added to guarantee the flow stability even for the case of the flow between walls. 

\subsection{Vlasov-like formulation}
Governing equations~\eqref{eq:VSW} admit a kinetic formulation~\cite{Zakharov} in the case of flows with non-zero vorticity (we choose $u_z>0$ as before). Following~\cite{TRC} we make a change of variables to new independent $t$, $x$, $u$ and dependent $W=1/u_z$, $u_b(t,x)=u(t,x,0)$, $u_s(t,x)=u(t,x,h)$ ones. For the unknown functions $W(t,x,u)$, $u_b(t,x)$ and $u_s(t,x)$ we obtain a closed integrodifferential model
\begin{equation} \label{eq:model-kin}
  \begin{array}{l}\displaystyle
	W_t+uW_x-gh_x W_u=0, \quad h=\int_{u_b}^{u_s} W\,du, \\[3mm]\displaystyle
	u_{bt}+u_b u_{bx}+gh_x=0,\quad u_{st}+u_s u_{sx}+gh_x=0,
  \end{array}
\end{equation}
which is analogous to the Vlasov kinetic equation. The variable $W$, which is inversely proportional to the long-wave vorticity, acts as a distribution function.

Indeed, due to the identities 
\[ du=u_t\,dt+u_x\,dx+u_z\,dz, \quad dz=z_t\,dt+z_x\,dx+z_u\,du \]
we have
\begin{equation} \label{eq:change-1} 
  u_t=-z_t u_z, \quad u_x=-z_x u_z, \quad u_z z_u=1. 
\end{equation}
Substituting relations~\eqref{eq:change-1} into the first equation~\eqref{eq:VSW} one obtains 
\[ z_t+uz_x-w-gh_xW=0, \]
where $W=z_u$. Further, we differentiate the above equation with respect to $u$ and take into account that $z_x-w_u=0$ (this formula is a direct consequence of the definition of $w$ following from~\eqref{eq:VSW}). As a result, we obtain the first equation of system~\eqref{eq:model-kin}. 

As a consequence of Eqs.~\eqref{eq:VSW} we obtain
\[ u_{jt}+u_j u_{jx}+gh_x=0,\quad (j=b,s). \]
While the equation for the velocity $u_b$ at the bottom $y=0$ is direct, the second equation for velocity $u_s=u$ at the free surface $z=h$ is less obvious. It can be obtained as follows:
\[ \begin{array}{l}\displaystyle 
     u_{st}+ u_s u_{sx}+gh_x= u_t+uu_x+u_z(h_t+uh_x)\big|_{z=h} +gh_x= \\[3mm]
     \quad\quad =u_t+uu_x+wu_z+gh_x \big|_{z=h}=0.   
   \end{array} \]
The second equation in \eqref{eq:VSW} reads
\[ \bigg(\int_{u_b}^{u_s} W\,du\bigg)_t + \bigg(\int_{u_b}^{u_s} uW\,du\bigg)_x =0. \]
It is easy to verify that this equation is fulfilled by virtue of \eqref{eq:model-kin}. 
\vspace{1mm}

{\sf Remark 2.} Kinetic formulation~\eqref{eq:model-kin} of the Benney equations~\eqref{eq:VSW} allows one to reveal an analogy between the stability criteria  for  plasma waves and shear flows. It is known~\cite{Stix} that any solution of the one-dimensional linearised Vlasov equation is stable if it is defined by a distribution function with a single maximum. Obviously, the functions $W=W(u)$ with one maximum obey the inequality
\[ (u_c-u)W'(u)\geq 0, \]
where $u=u_c$ is the extremum (maximum) point. Since the velocity profile $u=U(z)$ is related to the distribution function $W=W(u)$ as $u_z=1/W$ and, consequently, $W_u=-u_{zz}/u_z^3$, we obtain the Fjortoft stability criterion~\eqref{eq:Fjortoft}. 

\section{Class of piecewise linear velocity profiles: governing equations} 
\label{sec:piecewise-linear}

\begin{figure}[tbp]
 \begin{center}
  \resizebox{1\textwidth}{!}{\includegraphics{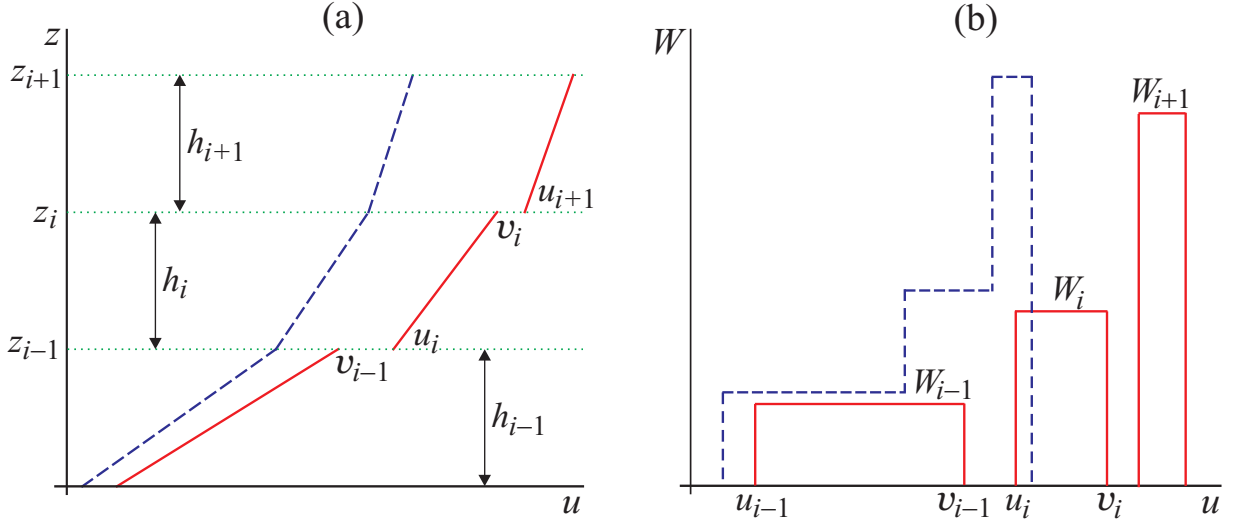}} \\[0pt]
 \end{center}
{\caption{An example of piecewise linear velocity profile (the variables $t$ and $x$ are fixed) {\rm (a)} and corresponding ``kinetic'' anzats {\rm (b)}; solid lines --- general case, dashed --- flow with piecewise constant vorticity and a continuous velocity profile.} \label{fig:pw-linear}} 
\end{figure}

Let us consider the class of flows with a piecewise linear velocity profile (see  Figure~\ref{fig:pw-linear})
\begin{equation} \label{eq:pw-lin-profile}  
  u=\omega_i(z-z_{i-1})+u_i, \quad z\in (z_{i-1},z_i), \quad i=1,...,N 
\end{equation}
and introduce the following notations. Each $i$-th layer is characterised by the velocity $u_i(t,x)$ at the lower boundary and the depth $h_i(t,x)=z_i-z_{i-1}$, as well as by the constant vorticity $\omega_i$. We also introduce velocity $v_i(t,x)=\omega_i h_i+u_i$ at the upper boundary of the layer and the average velocity $\bar{u}_i(t,x)=(v_i+u_i)/2=u_i+\omega_i h_i/2$. Obviously, that $z_0=0$, $z_N=h=h_1+...+h_N$. We also note that due to  the definition of $v_i$ we have
\begin{equation} \label{eq:vi-ui-hi-lin}  
  v_{it}+ v_i v_{ix}+ gh_x= u_{it}+ u_i u_{ix}+ gh_x + 
  \omega_i\big(h_{it}+(\bar{u}_i h_i)_x\big), \quad i=1,...,N. 
\end{equation}

At the upper and lower boundaries of each layer the following kinematic conditions should be satisfied 
\begin{equation} \label{eq:kin-cond} 
  \frac{\partial z_i}{\partial t}+v_i\frac{\partial z_i}{\partial x}=w_i^+, \quad 
  \frac{\partial z_{i-1}}{\partial t}+u_i\frac{\partial z_{i-1}}{\partial x}=w_i^-, 
  \quad i=1,...,N.
\end{equation}
Here $w_i^+$ and $w_i^-$ are the values of the vertical velocity $w$ at $z=z_i-0$ and $z=z_{i-1}+0$, correspondingly. Taking into account that representation~\eqref{eq:pw-lin-profile}, equation $u_x+w_z=0$ and kinematic condition~\eqref{eq:kin-cond} at $z=z_i-0$ one can express velocity $w$ in the form
\begin{equation} \label{eq:w-lin}  
  w=-(z-z_{i-1}) \bigg(\frac{\partial u_i}{\partial x}- 
  \omega_i\frac{\partial z_{i-1}}{\partial x}\bigg)+
  \frac{\partial z_{i-1}}{\partial t}+u_i\frac{\partial z_{i-1}}{\partial x}, 
  \quad z\in (z_{i-1},z_i).
\end{equation}
Let us calculate the difference $w_i^+-w_i^-$ using Eqs.~\eqref{eq:kin-cond} and   formula~\eqref{eq:w-lin}. As a result, we obtain
\begin{equation} \label{eq:hi-lin}  
  h_{it}+ \big(\bar{u}_i h_i\big)_x =0, \quad i=1,...,N. 
\end{equation}
Substitution of the velocities $u$ and $w$ given by formulae~\eqref{eq:pw-lin-profile}, \eqref{eq:w-lin} into the first equation of \eqref{eq:VSW} yields 
\begin{equation} \label{eq:ui-lin}  
  u_{it}+ u_i u_{ix}+ gh_x =0, \quad i=1,...,N. 
\end{equation}
Equations~\eqref{eq:hi-lin} and \eqref{eq:ui-lin} form a closed system for $2N$ unknown functions $h_i$ and $u_i$. In view of \eqref{eq:vi-ui-hi-lin} we can formulate the system governing flows from class~\eqref{eq:pw-lin-profile} in terms of $h_i$ and $v_i$. If all $\omega_i\neq 0$, then one can also use the variables $u_i$ and $v_i$.
\vspace{1mm}

{\sf Remark 3.} In terms of the Vlasov-like formulation~\eqref{eq:model-kin} the class of solutions~\eqref{eq:pw-lin-profile} corresponds to a piecewise constant distribution function $W$ in the form \cite{Ches_Pav}
\begin{equation} \label{eq:red-1}  
  W(t,x,u)=\sum\limits_{i=1}^N \Big(\theta(u-u_i(t,x))-\theta(u-v_i(t,x))\Big)W_i, 
\end{equation}
where $\theta$ is the Heaviside step-function, and $W_i=1/\omega_i$ are positive constants. The functions $u_i$ and $v_i$ are ordered in such a way that $u_{i+1}>v_i$ (see Figure~\ref{fig:pw-linear}). Substitution of \eqref{eq:red-1} into the first equation in \eqref{eq:model-kin} yields 
\[ \sum\limits_{i=1}^N \Big(\big(v_{it}+v_i v_{ix}+gh_x\big)\delta(u-v_i)-
   \big(u_{it}+u_i u_{ix}+gh_x\big)\delta(u-u_i)\Big)W_i =0, \]
where $\delta$ is the Dirac delta-function. Thus,  we obtain the following system of $2N$ PDEs  
\begin{equation} \label{eq:eq-2N}  
 \begin{array}{l}\displaystyle
  v_{it}+v_i v_{ix}+ gh_x=0, \quad  u_{it}+u_i u_{ix}+ gh_x=0, 
  \quad i=1,...,N \\[2mm]\displaystyle
  h=\sum\limits_{j=1}^N (v_j-u_j)/\omega_j
 \end{array}
\end{equation}
describing the piecewise constant anzats~\eqref{eq:red-1} of the Vlasov-like model~\eqref{eq:model-kin}, which corresponds to a free surface shear flow with a piecewise linear velocity profile. 
Note that   $u_b=u_1$ and  $u_s=v_N$ so the second and third equations in \eqref{eq:model-kin} are already included in \eqref{eq:eq-2N}. Obviously, that Eqs.~\eqref{eq:eq-2N} are equivalent to \eqref{eq:hi-lin}, \eqref{eq:ui-lin} if $\omega_i\neq 0$. System~\eqref{eq:eq-2N} represents the so-called {\it waterbag reduction} of the Benney equations. This type of reductions of kinetic equations play important role in plasma physics~\cite{davidson}. We note that system equivalent to \eqref{eq:eq-2N} appears in \cite{Konop_Bogdan} as a formal reduction of the dispersionless Kadomtsev--Petviashvili equation, outside any connection with vortical flows.
\vspace{1mm}

As it was mentioned above, shallow water equations for shear flows~\eqref{eq:VSW} can be rewritten in terms of the Riemann invariants~\eqref{eq:Riemann-gen} if the hyperbolicity condition~\eqref{eq:hyp-cond} is satisfied. In particular, for a piecewise linear velocity profile the Riemann invariants defined by formulae~\eqref{eq:Riemann-inv} are (see also~\cite{Konop_Bogdan})
\begin{equation} \label{eq:linear_Riemann_inv}  
 \begin{array}{l}\displaystyle
  r^i=k^i -\sum\limits_{j=1}^{N}\frac{g}{\omega_j}{\rm ln}\Big|\frac{v_j-k^i}{u_j-k^i}\Big|,
 \end{array}
\end{equation}
where $k^i(t,x)$ are zeros of the characteristic function~\eqref{eq:chi}, i.e. the roots of the equation
\begin{equation}\label{chi_k}
1-g\sum\limits_{i=1}^N \frac{1}{\omega_i} 
   \bigg(\frac{1}{v_i-k}-\frac{1}{u_i-k}\bigg)=0\,. 
 \end{equation}  
If system~\eqref{eq:eq-2N} is hyperbolic, it can be written in the form 
\begin{equation} \label{eq:r-2N} 
  r^i_t+k^i r^i_x =0, \quad\quad i=1,...,2N.
\end{equation}
Representation~\eqref{eq:r-2N}, in particular, allows one to construct solutions in the class of simple waves. The $m$-th family ($m=1, \dots, 2N$) of simple waves satisfies the relations
\[ r^i(u_1,...,u_N,v_1,...,v_N)=r^i_0={\rm const}, \quad i\neq m, 
   \quad k^m(u_1,...,u_N,v_1,...,v_N)=k(t,x), \] 
where $k(t,x)$ is a solution of the Hopf equation $k_t+kk_x=0$. {}

In some cases it is convenient to rewrite system~\eqref{eq:hi-lin}, \eqref{eq:ui-lin} in the form
\begin{equation} \label{eq:system-2N}  
 \begin{array}{l}\displaystyle
  h_{it}+\big(\bar{u}_i h_i \big)_x=0, \quad 
  \bar{u}_{it}+ \bigg(\frac{\bar{u}_i^2}{2}+\frac{\omega_i^2 h_i^2}{8}+
  g\sum\limits_{j=1}^N h_j\bigg)_x=0, \quad i=1,...,N.
 \end{array}
\end{equation}
One can see that in the limit of zero vorticity, $\omega_i \to 0$, $i=1,\dots, N$, system~\eqref{eq:system-2N} yields the Zakharov reduction of the Benney equations~\cite{Zakharov}. Importantly, as we will show in the next section, the presence in \eqref{eq:system-2N} of the terms related to vorticity enables one to describe multilayer flows with physically relevant, {\it continuous}, velocity profiles, not captured by the Zakharov reduction. 

It should be noted that system \eqref{eq:system-2N} admits a canonical Hamiltonian formulation:
\begin{equation}\label{ham1}
 \frac{\partial h_i}{\partial t}= -\frac{\partial}{\partial x}
   \bigg(\frac{\partial H}{\partial \bar{u}_i}\bigg), \quad
   \frac{\partial \bar{u}_i}{\partial t}= -\frac{\partial}{\partial x}
   \bigg(\frac{\partial H}{\partial h_i}\bigg), \quad i=1,...,N, 
   \end{equation}
where the Hamiltonian  is 
\begin{equation}\label{ham2}
H=\frac{1}{2}\sum\limits_{j=1}^N \bar{u}_j^2 h_j+ 
   \frac{g}{2}\bigg(\sum\limits_{j=1}^N h_j\bigg)^2+
   \frac{1}{24}\sum\limits_{j=1}^N \omega_j^2 h_j^3. \quad 
 \end{equation}
Equations~\eqref{eq:system-2N} obviously admit the energy and  momentum conservation laws, with the densities $H$ defined above and $P$ given below:
\[ P=\sum_{j=1}^{N}\bar{u}_{j}h_{j}. \]

Apart from the Hamiltonian structure the quasilinear system~\eqref{eq:system-2N} has a number of remarkable properties including the availability of infinitely many conservation laws and integrability. However, in this paper we focus only on the stability study leaving other  aspects related to mathematical properties of \eqref{eq:system-2N} for a separate publication.

\subsection{Class of piecewise linear continuous velocity profiles: hyperbolicity study} 
From system \eqref{eq:eq-2N} one can derive that the variables $s_i=u_{i+1}-v_i$ satisfy the equations 
\[ s_{it}+\Big(\frac{u_{i+1}+v_i}{2}s_i\Big)_x=0. \] 
Obviously, if $s_i|_{t=0}=0$ then $s_i=0$ for all $t>0$. This follows from the uniqueness of the solution of the Cauchy problem for the above system. Thus, for a homogeneous fluid, if the initial velocity profile is continuous, it will stay continuous for all time. For the density-stratified fluid this statement is not valid. 

Sliding between the layers is unusual for homogeneous fluids. Therefore, from the physical point of view is more natural to consider flows with a continuous velocity profile. $N$-layer flows of homogeneous fluid with a piecewise constant vorticity and a continuous velocity profile are defined by \eqref{eq:pw-lin-profile}, where $u_i=v_{i-1}$. In this case system~\eqref{eq:hi-lin}, \eqref{eq:ui-lin} takes the form
\begin{equation} \label{eq:red-cont}  
 \begin{array}{l}\displaystyle
  h_{it}+(\bar{u}_i h_i)_x=0, \quad  i=1,...,N \\[2mm]\displaystyle 
  v_{0t}+v_0 v_{0x}+gh_x=0, \quad h=\sum\limits_{i=1}^N h_i
 \end{array}
\end{equation}
and consists of $N+1$ equations for the depths $h_i$ and for the velocity $v_0$ at the bottom $z=0$. Here 
\begin{equation}\label{eq:27}
 \bar{u}_i=v_i -\frac{\omega_i h_i}{2}, \quad v_i=v_0+\sum\limits_{j=1}^i \omega_j h_j \,. 
 \end{equation}

The Riemann invariants for \eqref{eq:red-cont} are obtained from the general formulae~\eqref{eq:linear_Riemann_inv} by the reduction $u_j=v_{j-1}$. We stress that system~\eqref{eq:red-cont}, \eqref{eq:27}  represents an integrable multilayer approximation of the Benney equations, which, unlike the Zakharov reduction~\cite{Zakharov}, describes flows with  continuous velocity profiles. As a matter of fact, the presence of piecewise-constant vorticity plays the crucial role in our construction. 

Now, using the generalized theory of characteristics~\cite{Tesh85,LT00} introduced before, we formulate sufficient conditions for hyperbolicity of Eqs.~\eqref{eq:red-cont}. 
System~\eqref{eq:red-cont} can be written in the form 
\begin{equation} \label{eq:system-vector-form}  
  \mathbf{u}_t+A(\mathbf{u})\mathbf{u}_x=0, 
\end{equation}
where $\mathbf{u}=(h_1,...,h_N,v_0)^{\rm T}$ is the unknown vector and $A(\mathbf{u})$ is the corresponding matrix. To find the eigenvalues of $A(\mathbf{u})$, one have to solve the equation 
\begin{equation} \label{eq:det} 
  D(k)=\det(A-kI)=0. 
\end{equation}
It is convenient to use the characteristic function $\chi(k)$ \eqref{eq:chi}. We introduce the separate notation $\bar{\chi}(k)$   for this function evaluated on the piecewise-linear velocity profile~\eqref{eq:pw-lin-profile} with the additional requirement of continuity $u_i=v_{i-1}$. Then, on using the second formula in \eqref{eq:27} we obtain, 
\[ \begin{array}{l}\displaystyle
    \bar{\chi}(k)=1+ g\sum\limits_{i=1}^N \frac{1}{\omega_i}
    \bigg(\frac{1}{v_i-k}-\frac{1}{v_{i-1}-k} \bigg)= 
    1-\sum\limits_{i=1}^N \frac{gh_i}{(v_i-k)(v_{i-1}-k)} = \\[3mm]\displaystyle 
    \quad\quad\quad
    =1-\frac{g}{\omega_1}\frac{1}{v_0-k}+\frac{g}{\omega_N}\frac{1}{v_N-k}-
    g\sum_{i=1}^{N-1}\bigg(\frac{1}{\omega_{i+1}}-\frac{1}{\omega_i}\bigg)\frac{1}{v_i-k}\,.
  \end{array} \]      
The roots $k=k^i$ of the equation $\bar{\chi}(k)=0$ are the characteristic velocities of system~\eqref{eq:red-cont}, and there is the following relation between the polynomial $D(k)$ and $\bar{\chi}(k)$
\begin{equation} \label{eq:det-chi}  
  D(k)=\bar{\chi}(k)\prod\limits_{j=0}^N (v_j-k). 
\end{equation}
The derivative of the function $\bar{\chi}(k)$ is 
\begin{equation}\label{eq:31}
 \begin{array}{l}\displaystyle
    \bar{\chi}'(k)=g\sum\limits_{i=1}^N \frac{1}{\omega_i}
        \bigg(\frac{1}{(v_i-k)^2}-\frac{1}{(v_{i-1}-k)^2}\bigg)\,.
   \end{array} 
   \end{equation}

{\sf Lemma 4.} {\it Let all constant vorticities $\omega_i$ be ordered  
\begin{equation}  \label{eq:Rayleigh}
  \omega_{N}<\omega_{N-1}<...<\omega_1 \quad {\rm or} \quad \omega_{N}>\omega_{N-1}>...>\omega_1.
\end{equation}
Then system~\eqref{eq:red-cont} is hyperbolic. }
\vspace{1mm}

\begin{figure}[tbp]
 \begin{center}
  \resizebox{0.6\textwidth}{!}{\includegraphics{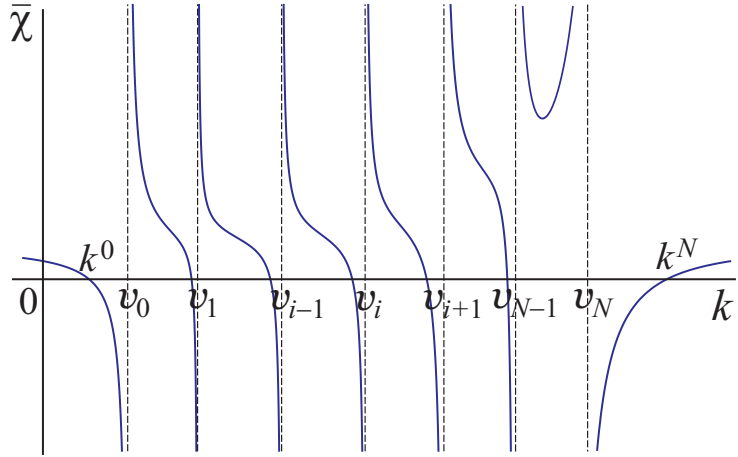}} \\[0pt]
 \end{center}
{\caption{A typical graph of the function $\bar{\chi}(k)$ for the case $\omega_1>\omega_2>...>\omega_N>0$.}\label{fig:pw-lin-chi}} 
\end{figure}

{\sf Proof of Lemma 4.} Let us suppose that all $\omega_i$ have the same sign $\omega_i>0$. We prove this statement for the case $\omega_1>...>\omega_N$ (the other cases are treated similarly). In the intervals $k \in (-\infty, v_0)$ and $k \in (v_N, \infty)$ the equation $\bar{\chi}(k)=0$ has exactly two real roots $k^0<v_0$ and $k^N>v_N$ (see Figure~\ref{fig:pw-lin-chi}). Indeed, $\bar{\chi}(k)\to 1$ if $k\to\pm\infty$, $\bar{\chi}(k)\to-\infty$ if $k\to v_0-0$, and $\bar{\chi}(k)\to\infty$ if $k\to v_N+0$. Moreover, $\bar{\chi}(k)'<0$ if $k\in (-\infty, v_0)$, and $\bar{\chi}(k)'>0$ if $k\in (v_N,\infty)$, see~\eqref{eq:31}. The function $\bar{\chi}(k)$ is continuous on the intervals $k\in (v_i,v_{i+1})$ and has the following limiting values 
\[ \lim\limits_{k\to v_i -0} \bar{\chi}(k)=-\infty, \quad 
   \lim\limits_{k\to v_i +0} \bar{\chi}(k)= \infty, \quad i=0,...,N-1. \]  
Thus, function $\bar{\chi}(k)$ change sign in the intervals $k\in (v_{i-1},v_i)$, $i=1,...,N-1$ and the equation $\bar{\chi}(k)=0$ has (at least) one root $k=k^i$ on each of these intervals (see Figure~\ref{fig:pw-lin-chi}). We show that the function $\bar{\chi}(k)$ has at least $N+1$ zeros $k=k^i$, $k^i\neq v_j$. According to definition~\eqref{eq:det}, $D(k)$ is a polynomial of order $N+1$ and, consequently, has $N+1$ roots. Taking into account relation~\eqref{eq:det-chi} one can conclude that the function $\bar{\chi}(k)$ has exactly $N+1$ zeros $k=k^i\neq v_j$. 

Let the vorticities $\omega_i$ are ordered and change sign such that
\[ \omega_1>...>\omega_j>0>\omega_{j+1}>...>\omega_N. \]
Since $\omega_{i+1}^{-1}-\omega_i^{-1}>0$ for all $i$ except for $i=j$, the function $\bar{\chi}(k)$ has the following limiting values 
\[ \lim\limits_{k\to v_i -0} \bar{\chi}(k)=-\infty, \quad 
   \lim\limits_{k\to v_i +0} \bar{\chi}(k)= \infty, \quad i=0,...,j-1,j+1,...,N. \]  
Suppose that  all the characteristic velocities $v_k$ are pairwise distinct, i.e. $v_i\neq v_m$ for $i\neq m$  and thus can be ordered. To this end, instead of $v_i$ we  introduce the variables $q_i$ ($q_l=v_m$) which are ordered such that $q_0<q_1<...<q_N=v_j$. In this case Figure~\ref{fig:pw-lin-chi} (where $q_i$ stand for $v_i$) also represents a typical graph of the function $\bar{\chi}(k)$ having $N+1$ real roots $k^i\neq v_m$. 
We also note, that if $v_l=v_m$ ($l\leq j, m>j$), then $k=v_l$ is a root of the equation $D(k)=0$. Hence, the cases of coinciding velocities ($v_l=v_m$) and zero vorticity ($\omega_j=0$) are also included into consideration. 

Thus, we proved that conditions~\eqref{eq:Rayleigh} provide the existence of $N+1$ different characteristic roots $k=k^i$ of Eq.~\eqref{eq:det}. This means that system~\eqref{eq:red-cont} is hyperbolic. 
\vspace{1mm}

Inequalities~\eqref{eq:Rayleigh} imply that system~\eqref{eq:red-cont} is strictly hyperbolic and hence the flow is stable in the sense of well-posedness of time evolution, see~\cite{Chumakova_etal1, Chumakova_etal2}. This sufficient condition~\eqref{eq:Rayleigh} is reminiscent of the famous Rayleigh stability criterion about the shear flow stability between rigid walls: if the velocity profile is convex, the flow is stable. It weakens the criterion of stability proven in Subsection 2.1 to the case of piecewise linear velocity profiles.
\vspace{1mm}

{\sf Remark 4.} Eqs.~\eqref{eq:red-cont} for two-layer flows ($N=2$) are always hyperbolic because the two-layer velocity profile is always convex. Indeed, equation $\bar{\chi}(k)=0$ has two real roots $k^l<\min v_j$ and $k^r>\max v_j$ ($j=0,1,2$). Therefore, polynomial $D(k)$ has three real zeroes. 
\vspace{1mm}

{\sf Remark 5.} For multilayer flows the violation of conditions~\eqref{eq:Rayleigh} may lead to the loss of hyperbolicity of Eqs.~\eqref{eq:red-cont}. Let us consider the following example of a three-layer flow ($N=3$) with unordered positive vorticities ($\omega_1<\omega_2$, $\omega_3<\omega_2$). These parameters correspond to the piecewise linear approximation of smooth non-convex velocity profile of type~\eqref{eq:class-st-unst} (see Figure~\ref{fig:poss-unst}) when Fjortoft-like criterion~\eqref{eq:Fjortoft} can not be applied. We choose $g=1$, $h_1=h_3=1$, $h_2=\alpha>0$, and $\omega_1=1/2$, $\omega_2=1$, $\omega_3=1/4$. It is easy to verify that there are four real roots of the characteristic equation $\bar{\chi}(k)=0$ if $\alpha>\alpha_*\approx 0.885$. For $\alpha<\alpha_*$, there are only two real roots of the equation. Hence, system~\eqref{eq:red-cont} is not hyperbolic in this case. 

Let us choose positive constants $\omega_i>0$ such that $\omega_1>\omega_2$, $\omega_3>\omega_2$. It corresponds to three-layer ($N=3$) piecewise linear approximation of Fjortoft-like velocity profile (see Figure~\ref{fig:Fjortoft}). In this case system~\eqref{eq:red-cont} is always hyperbolic, because equation $\bar{\chi}(k)=0$ has four real roots: $k^0<v_0$, $k^1\in (v_0,v_1)$, $k^2\in (v_2,v_3)$, and $k^3>v_3$. Indeed, $\omega_2^{-1}-\omega_1^{-1}>0$ and $\omega_3^{-1}-\omega_2^{-1}<0$. Hence, $\bar{\chi}(k)\to +\infty$ as $k\to v_0+0$ and $k\to v_3-0$; $\bar{\chi}(k)\to -\infty$ as $k\to v_1-0$ and $k\to v_2+0$. This means that there are roots $k^1$ and $k^2$ on the intervals $(v_0,v_1)$ and $(v_2,v_3)$, correspondingly. 

\section{Two-layer stratified flow with a piecewise constant vorticity}  
The generalization of the above results to the case of multilayer stratified flows is less obvious. Indeed, the fact that the densities in each layer are different implies that even if the sliding at the fluid interfaces was vanishing initially, it can appear during the evolution. So, a continuous velocity profile does not exist in this case. The remarkable fact of the existence of Riemann invariants for homogeneous multilayer system is also absent for stratified $N$-layer flows. Indeed, our calculation of the Haantjes tensor (see Appendix) shows that it vanishes identically only in the case of homogeneous fluids. 

\begin{figure}[tbp]
 \begin{center} 			
  \resizebox{.6\textwidth}{!}{\includegraphics{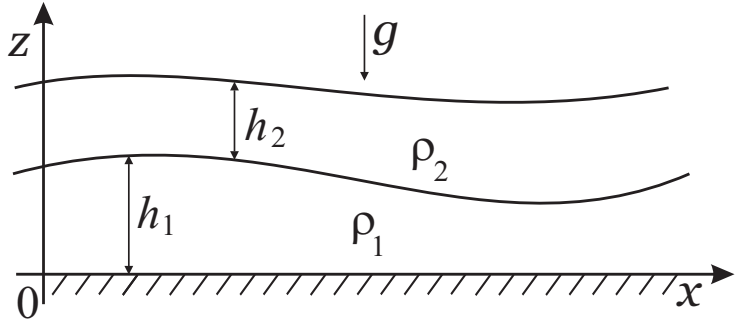}} \\[0pt]
  \caption{Schematic of the two-layer stratified flow.} \label{fig:fig_sketch}
 \end{center} 
\end{figure}

We present here the hyperbolicity analysis for two-layer stratified flows. A general two-layer system is composed of two immiscible fluids of different constant densities $\rho_1$ and $\rho_2$ confined between the upper free surface and the lower rigid boundary. The shear flow in the long-wave approximation is governed by the equations~\cite{chesn04}
\begin{equation} \label{eq:2l-longwaves}
  \begin{array}{l} \displaystyle
    u^1_t+ u^1 u^1_x+ w^1 u^1_z+ gh_{1x}+ g\rho h_{2x}=0, \quad 
    h_{1t}+\Big(\int\limits_0^{h_1}u^1\,dz\Big)_x=0, \\[3mm]\displaystyle 
    u^2_t+u^2 u^2_x+w^2 u^2_z+ gh_{1x}+ gh_{2x}=0, \quad 
    h_{2t}+\Big(\int\limits_{h_1}^{h_1+h_2}u^2\,dz\Big)_x=0, \\[3mm]\displaystyle 
    w^1=-\int\limits_0^z u^1_x(t,x,z')\,dz', \quad 
    w^2=-\int\limits_{h_1}^z u^2_x(t,x,z')\,dz'+h_{1t}+u^2(t,x,h_1)h_{1x}.
 \end{array}
\end{equation}
Here the variables $u^i(t,x,z)$, $w^i(t,x,z)$ and $h_i(t,x)$ are the velocity components and the layer depths; $g$ is the gravity acceleration and $\rho\leq 1$ is a parameter defined by $\rho=\rho_2/\rho_1$. The subscript $i=1$ and $2$ corresponds to the lower and upper layers of the fluid respectively (see Figure~\ref{fig:fig_sketch}). 

It should be noted that in the approximation considered, the vorticity in the layer is proportional to $u^i_z$, and in the case of no velocity shear, system~\eqref{eq:2l-longwaves} reduces to the well-known equations of two-layer shallow water~\cite{ovs79}.

Eqs.~\eqref{eq:2l-longwaves} describing two-layer shear flows were studied in  \cite{chesn04} where characteristic function was obtained in the form
\begin{equation} \label{eq:char2}
 \begin{array}{l}\displaystyle
  \hat{\chi}(k)=1-g\int\limits_0^{h_1} \frac{dz}{(u^1-k)^2}- g\int\limits_{h_1}^{h_1+h_2} \frac{dz}{(u^2-k)^2} + \\[3mm]\displaystyle
  \quad\quad\quad\quad\quad 
  +(1-\rho)g^2 \int\limits_0^{h_1} \frac{dz}{(u^1-k)^2} \int\limits_{h_1}^{h_1+h_2} \frac{dz}{(u^2-k)^2}.
  \end{array}
\end{equation}
Equation $\hat{\chi}(k)=0$ defines the velocity of perturbation propagation in the fluid. In the case of stratified fluid ($\rho<1$) the characteristic function $\hat{\chi}(k)$ involves nonlinear term (with multiplication of integrals of the functions $1/(u^i-k)^2$ over the depths of the lower and upper layers). This complicates the analysis and formulation of the hyperbolicity conditions for Eqs.~\eqref{eq:2l-longwaves}. 

\begin{figure}[tbp]
 \begin{center} 			
  \resizebox{.9\textwidth}{!}{\includegraphics{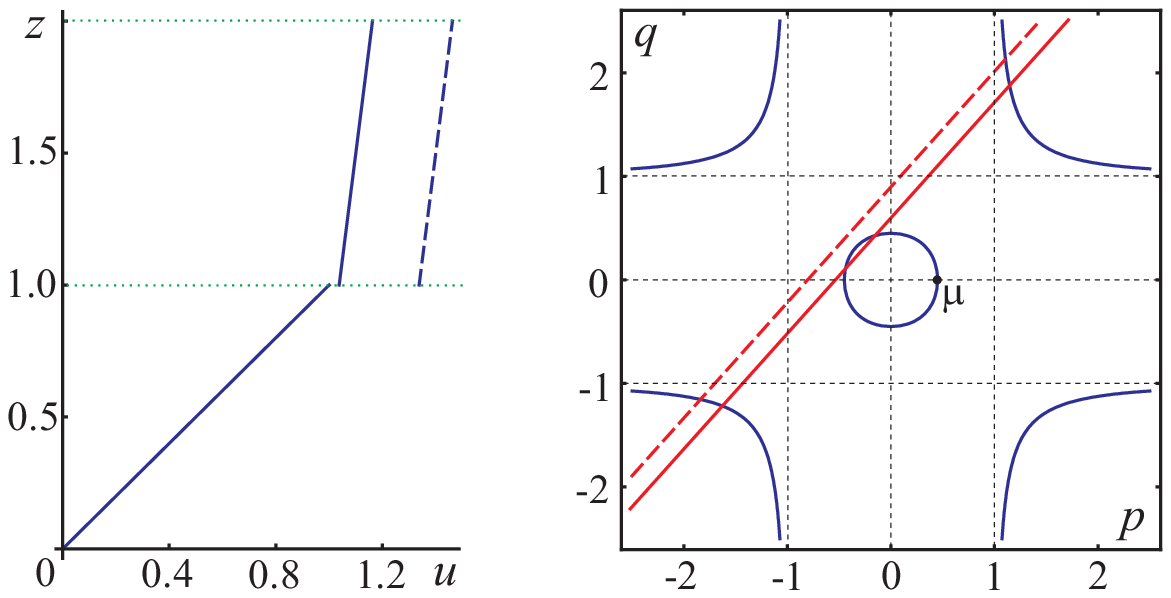}} \\[0pt]
 \end{center} 
\parbox{0.48\textwidth}{\caption{Piecewise linear velocity profile~\eqref{eq:lin-vel} for $h_1=h_2=1$, $\omega_1=1$, $\omega_2=1/8$, $\bar{u}_1=\omega_1 h_1/2$, $\bar{u}_2=\bar{u}_1+0.6$ (solid) and $\bar{u}_2=\bar{u}_1+0.9$ (dashed line).} \label{fig:stab_lin_vel-a}} \hfill
\parbox{0.48\textwidth}{\caption{The curve~\eqref{eq:curve} and the straight lines~\eqref{eq:line} in $(p,q)$-plane for the same parameters as in \ref{fig:stab_lin_vel-a} and $g=1$, $\rho=1$.} \label{fig:stab_lin_vel-b}}
\end{figure}

Let us consider the following class of flows
\begin{equation} \label{eq:lin-vel}
   \begin{array}{ll} \displaystyle
     u^1(t,x,z)=\omega_1 z+u_1= \omega_1\Big(z-\frac{h_1}{2}\Big)+\bar{u}_1,  & z \in (0,h_1), \\[2mm]\displaystyle
     u^2(t,x,z)=\omega_2(z-h_1)+u_2= \omega_2\Big(z-h_1-\frac{h_2}{2}\Big)+\bar{u}_2,  & z \in (h_1,h_1+h_2),
   \end{array}
\end{equation}
where as before $\omega_i$ ($i=1,2$) are the constant vorticities in the layers, $u_i$ are the velocities at the lower boundaries of the layers (at $z=0$ and $z=h_1+0$), and $\bar{u}_i(t,x)$ are the layer-averaged velocities. The corresponding velocity profile is presented in Figure~\ref{fig:stab_lin_vel-a} for $\omega_1=1$, $\omega_2=1/8$, $h_1=h_2=1$, $\bar{u}_1=\omega_1 h_1/2$, $\bar{u}_2=\bar{u}_1+0.6$ (solid) and $\bar{u}_2=\bar{u}_1+0.9$ (dashed line). These profiles differ only in the magnitude of the sliding at the fluid interface. 

In this case Eqs.~\eqref{eq:2l-longwaves} take the form
\begin{equation} \label{eq:2l-lin_vel}
   \begin{array}{l} \displaystyle
      \bar{u}_{1t}+\bar{u}_1 \bar{u}_{1x}+\Big(g+\frac{\omega_1^2 h_1}{4}\Big)h_{1x}+g\rho h_{2x}=0, \quad 
      h_{1t}+(h_1\bar{u}_1)_x=0, \\[3mm]\displaystyle 
     \bar{u}_{2t}+\bar{u}_2 \bar{u}_{2x}+gh_{1x}+\Big(g+\frac{\omega_2^2 h_2}{4}\Big)h_{2x}=0, \quad 
     h_{2t}+(h_2\bar{u}_2)_x=0.
   \end{array}
\end{equation}
If $\rho=1$ (homogeneous fluid) Eqs.~\eqref{eq:2l-lin_vel} coincide with system~\eqref{eq:system-2N} for $N=2$. To study hyperbolicity of Eqs.~\eqref{eq:2l-lin_vel} we rewrite this system in form~\eqref{eq:system-vector-form}, where $\mathbf{u}=(h_1,h_2,\bar{u}_1,\bar{u}_2)^{\rm T}$ is the unknown vector, and $A(\mathbf{u})$
is a matrix of $4\times 4$. The eigenvalues of $A(\mathbf{u})$ are determined by equations
\begin{equation} \label{eq:char1} 
   D(k)=\big((\bar{u}_1-k)^2-\alpha_1 h_1\big)\big((\bar{u}_2-k)^2-\alpha_2 h_2\big)- g^2\rho h_1 h_2=0,
\end{equation}
where
\[ \alpha_1=g+\frac{\omega_1^2 h_1}{2},  \quad 
   \alpha_2=g+\frac{\omega_2^2 h_2}{2}\,. \]
System~\eqref{eq:2l-lin_vel} is hyperbolic if equation~\eqref{eq:char1} has four real roots. 

The characteristic velocities $k$ can be directly obtained from equation $\hat{\chi}(k)=0$. Indeed, substituting piecewise linear velocity profile~\eqref{eq:lin-vel} in \eqref{eq:char2} leads to the following relation 
\[ D(k)=(u_1-k)(v_1-k)(u_2-k)(v_2-k\big)\hat{\chi}(k), \]
where $u_i$ and $v_i$ are the fluid velocities at the lower and upper boundaries of the layers. As was shown before, in particular case $u_2=v_1$ (continuous velocity profile) and $\rho=1$ (homogeneous fluid) the considered model is always hyperbolic. 

An insightful geometric interpretation of the characteristics proposed by Ovsyannikov~\cite{ovs79} for two-layer potential flows ($\omega_i=0$) can be applied here. We introduce the new variables $p$ and $q$ by the formulae
\begin{equation} \label{eq:pq} 
   \bar{u}_1-k=p\sqrt{\alpha_1 h_1}, \quad \bar{u}_2-k=q\sqrt{\alpha_2 h_2}. 
\end{equation}
Then equation \eqref{eq:char1} can be rewritten in the form
\begin{equation} \label{eq:curve}  
  (p^2-1)(q^2-1)=\frac{g^2\rho}{\alpha_1\alpha_2}\,.  
\end{equation}
In the $(p,q)$-plane equation~\eqref{eq:curve} describes a fourth-order curve with four symmetry axis (see Figure~\ref{fig:stab_lin_vel-b}). The variables $p$ and $q$ in virtue of \eqref{eq:pq} are related by
\begin{equation} \label{eq:line}  
   q=p\sqrt{\alpha_1/\alpha_2}+(\bar{u}_2-\bar{u}_1)/\sqrt{\alpha_2}.
\end{equation}
The number of the real roots of equation~\eqref{eq:char1} is determined be the number of intersections of the curve~\eqref{eq:curve} with straight-line~\eqref{eq:line}. It is clear that a necessary condition for the existence of 4 real roots is the following inequality
\[ \mu=\sqrt{1-\frac{g^2 \rho}{\alpha_1\alpha_2}}\,>0, \]
which is always fulfilled if $0<\rho\leq 1$ and $|\omega_1|+|\omega_2|>0$. In the case of potential flow ($\omega_i=0$) the ``radius'' $\mu$ is $\sqrt{1-\rho}$. Hence, the presence of vorticity improves stability of the two-layer flow. In particular, even for a homogeneous fluid ($\rho=1$) the flow can be stable, if the sliding between the fluid layers is sufficiently small (see Figures~\ref{fig:stab_lin_vel-a} and \ref{fig:stab_lin_vel-b}).

\section{Conclusion}
The classical stability criteria of shear flows (Rayleigh, Fjortoft) are typically obtained for flows between rigid walls. Stability of shear flows with free surface has been much less studied. The main goal of this work was to analyse stability of shallow shear flows with free surface in terms of hyperbolicity of the nonlinear governing equations. First, we outlined the general hyperbolicity conditions~\eqref{eq:hyp-cond} of the Benney equations~\eqref{eq:VSW} introduced by Teshukov~\cite{Tesh85, LT00}. Further, we have proved in Subsection 2.1 that the monotonicity and convexity of the velocity profile are sufficient for the stability of shallow water shear flows with a free surface. This result is also true for the Fjortoft-like velocity profiles~\eqref{eq:Fjortoft}. Moreover, we presented the class of flows~\eqref{eq:class-st-unst} for which the hyperbolicity conditions~\eqref{eq:hyp-cond} may be violated. Kinetic formulation~\eqref{eq:model-kin} of the governing equations allows one to show the analogy between the stability criteria for plasma waves and shear flows. 

In the subsequent sections we focus our attention on the multilayer flows with piecewise linear (discontinuous or continuous) velocity profile describing by models~\eqref{eq:eq-2N} and \eqref{eq:red-cont}. We have revealed some important mathematical properties of the models such as the existence of Riemann invariants~\eqref{eq:linear_Riemann_inv} and the Hamiltonian structure \eqref{ham1}, \eqref{ham2}. We have shown that the presence of non-zero vorticity enables one to find multilayer integrable reductions of the Benney system describing shear flows with a class of physically natural continuous velocity profiles, improving the properties of the well-known Zakharov's reductions. For the class of flows with piecewise linear continuous velocity profile we formulated sufficient conditions of stability~\eqref{eq:Rayleigh} which are reminiscent of the famous Rayleigh--Fjortoft criterion. The generalization of the results obtained for layered flows of homogeneous fluid to the case of density stratified flows is less obvious. In particular, a continuous velocity profile does not exist. Moreover, the Haantjes tensor~\eqref{eq:Haantjes} does not vanish for system~\eqref{eq:2l-lin_vel} if the density ratio $\rho\neq 1$. This mean that the system does not admit Riemann invariants. Nevertheless, we have been able to show that the presence of vorticity has stabilizing effect on the flow. 

\section*{Acknowledgements} This work was supported by the Russian Science Foundation (grant No. 15-11-20013). 

The authors thank IM\'{e}RA foundation of Aix-Marseille Universit\'{e} for hospitality. They also grateful to P.\,V.~Kovtunenko for performing symbolic computations in Appendix.

\section*{Appendix. The Haantjes tensor: the diagonalisability criterion.} Any strictly hyperbolic system of quasilinear equations of the type
\[ a_t^i+v_j^i(\mathbf{a})a_x^j=0, \quad i,j=1,...,M \]
can be diagonalised, i.e. can be rewritten in terms of Riemann invariants if and only if the Haantjes tensor \cite{Haantjes} constructed in terms of the matrix $v_j^i(\mathbf{a})$ is identically vanishing \cite{FerTsar}. For computing of Haantjes tensor one calculates first the Nienhuis tensor 
\[ N_{jk}^i=v_j^p\partial_p v_k^i- v_k^p \partial_p v_j^i- v_p^i(\partial_j v_k^p- 
   \partial_k v_j^p), \quad \partial_p\equiv \partial/\partial a^p, \]
and then finally the Haantjes tensor%
\begin{equation} \label{eq:Haantjes}
  H_{jk}^{i}=N_{pn}^{i}v_{j}^{p}v_{k}^{n}-N_{jn}^{p}v_{p}^{i}v_{k}^{n}-   
  N_{nk}^{p}v_{p}^{i}v_{j}^{n}+N_{jk}^{p}v_{n}^{i}v_{p}^{n}. 
\end{equation}
Symbolic computations show that the Haantjes tensor~\eqref{eq:Haantjes} vanishes identically for system~\eqref{eq:2l-lin_vel} if and only if $\rho=1$. This fact justifies the existence of Riemann invariants given explicitly by \eqref{eq:linear_Riemann_inv} for the case of homogeneous fluid.


\begin{thebibliography}{}

\bibitem{Stoker} 
{\sc J.~J. Stoker}, {\em Water Waves: The Mathematical Theory with Applications}, Interscience, New York, 1957.

\bibitem{Burns} 
{\sc J.~C. Burns}, {\em Long waves in running water}, Proc. Cambridge Philos. Soc. 49 (1953), pp.~695--706.

\bibitem{Benney} 
{\sc D.~J. Benney}, {\em Some properties of long nonlinear waves}, Stud. Appl. Math. 52  (1973), pp.~45--50.

\bibitem{Zakharov} 
{\sc V.~E. Zakharov}, {\em Benney equations and quasi-classical approximation in the method of inverse problem}, Funk. Anal. Prilozh. 14 (1980), pp.~15--24.

\bibitem{TRC} 
{\sc V. Teshukov, G. Russo, and A. Chesnokov}, {\em Analytical and numerical solutions of the shallow water equations for 2-D rotational flows}, Math. Models Methods Appl. Sci. 14 (2004), pp.~1451--1479.

\bibitem{KupMan} 
{\sc B.~A. Kupershmidt and Yu.~I. Manin}, {\em Equations of long waves with a free surface. II. Hamiltonian structure and higher equations}, Funk. Anal. Priloz. 12 (1978), pp.~20--29. 

\bibitem{LebMan}{\sc D.~R. Lebedev and Yu.~I. Manin}, {\em   Conservation laws and Lax representation of Benney's long wave equations} Phys. Lett. A, 74 (1979), pp.~154-156.

\bibitem{Freeman} 
{\sc N.~C. Freeman}, {\em Simple waves on shear flow: similarity solutions}, J. Fluid Mech. 56 (1972), pp.~257--263.

\bibitem{Sachdev} 
{\sc P.~L. Sachdev}, {\em Self-similarity and Beyond: Exact Solutions of Nonlinear Problems}, Boca Raton, FL: Chapman \& Hall/CRC, 2000.

\bibitem{Varley} 
{\sc E. Varley and P.~A. Blythe}, {\em Long eddies in sheared flows}, Stud. Appl. Math. 68 (1983), pp.~103--187.

\bibitem{Drazin} 
{\sc P.~G. Drazin} {\em Introduction to Hydrodynamic Stability}, Cambridge University Press, Cambridge, 2002.

\bibitem{Chumakova_etal1}
{\sc L. Chumakova, F.~E. Manzaque, P.~A. Milewski, R.~R. Rosales, E.~G. Tabak, and C.~V. Turner}, {\em Shear instability for stratified hydrostatic flows}, Comm. Pure Appl. Math. 62 (2009), pp.~183--197.

\bibitem{Chumakova_etal2}
{\sc L. Chumakova, F.~E. Manzaque, P.~A. Milewski, R.~R. Rosales, E.~G. Tabak, and C.~V. Turner}, {\em Stability properties and nonlinear mappings of two and three-layer stratified flows}, Stud. Appl. Math. 122 (2009), pp.~123--137.

\bibitem{Tesh85} 
{\sc V.~M. Teshukov}, {\em Hyperbolicity of long-wave equations}, Dokl. Akad. Nauk. 284 (1985), pp.~555--559.

\bibitem{Tesh94} 
{\sc V.~M. Teshukov}, {\em Long waves in an eddying barotropic liquid}, J. Appl. Mech. Tech. Phys. 35 (1994) pp.~823--831.

\bibitem{LT00} 
{\sc V.~Yu. Liapidevskii and V.~M. Teshukov}, {\em Mathematical Models of Propagation of Long Waves in a Non-Homogeneous Fluid}, Novosibirsk, Siberian Division of the Russian Academy of Sciences, 2000 (in Russian).

\bibitem{Chesn13}
{\sc A.~A. Chesnokov and A.~K. Khe}, {\em Is Landau Damping Possible in a Shear Fluid Flow?} Stud. Appl. Math. 131 (2013), pp.~343--358.

\bibitem{Tesh95} 
{\sc V.~M. Teshukov and M.~M. Sterkhova}, {\em Characteristic properties of the system of equations of a shear flow with non-monotonic velocity profile}, J. Appl. Mech. Tech. Phys. 36 (1995), pp.~367--372.

\bibitem{KnCn12} 
{\sc E.~Yu. Knyazeva and A.~A. Chesnokov}, {\em Stability criterion of shear fluid flow and the hyperbolicity of the long-wave equations}, J. Appl. Mech. Tech. Phys. 53 (2012), pp.~657--663. 

\bibitem{Gavrilyuk_Teshukov} 
{\sc S.~L. Gavrilyuk and V.~M. Teshukov}, {\em Linear stability of parallel inviscid flows of shallow water and bubbly fluid}, Stud. Appl. Math. 113 (2004), pp.~1--29.

\bibitem{Stix}
{\sc T. Stix}, {\em The Theory of Plasma Waves}, McGarw-Hill, New York, 1962.

\bibitem{Ches_Pav}
{\sc A.~A. Chesnokov and M.~V. Pavlov}, {\em Reductions of kinetic equations to finite-component systems}, Acta Appl. Math. 122 (2012), pp.~367--380.

\bibitem{davidson} 
{\sc R.~C.~Davidson}, {\em Methods in Nonlinear Plasma Theory}, Academic Press, New York and London, 1972.

\bibitem{Konop_Bogdan} {\sc L.~V. Bogdanov and B.~G. Konopelchenko}, {\em Symmetry constraints for dispersionless integrable equations and systems of hydrodynamic type}, Phys. Lett. A 330 (2004), pp.~448--459.

\bibitem{chesn04}
{\sc A.~A. Chesnokov}, {\em On the propagation of long-wave perturbations in a two-layer free-boundary rotational fluid}, J. Appl.Mech. Tech. Phys. 45 (2004), pp.~230--238.

\bibitem{ovs79}
{\sc L.~V. Ovsyannikov}, {\em Two-layer shallow-water model}, J. Appl.Mech. Tech. Phys. 20  (1979), pp.~127--135.  

\bibitem{Haantjes} 
{\sc J. Haantjes}, {\em On $X_{m}$--forming sets of eigenvectors}, {\em Indagationes Mathematicae} 17 (1955), pp.~158--162.

\bibitem{FerTsar} 
{\sc E.~V. Ferapontov and S.~P. Tsarev}, {\em Hydrodynamic type systems, arising in gas chromatography. Riemann invariants and exact solutions}, Math. Modelling, 3 (1991), pp.~82--91 (in Russian). 

\end{thebibliography}
\end{document}